\documentclass[lettersize,journal]{IEEEtran}
\usepackage{amsmath,amsfonts}
\usepackage{algorithmic}
\usepackage{algorithm}
\usepackage{array}
\usepackage[caption=false,font=normalsize,labelfont=sf,textfont=sf]{subfig}
\usepackage{textcomp}
\usepackage{stfloats}
\usepackage{url}
\usepackage{verbatim}
\usepackage{graphicx}
\usepackage{cite}
\usepackage{listings}
\usepackage{xcolor}

\definecolor{codegray}{gray}{0.95}

\lstdefinestyle{mypython}{
    language=Python,
    basicstyle=\ttfamily\footnotesize,
    keywordstyle=\color{blue},
    commentstyle=\color{gray},
    stringstyle=\color{red},
    backgroundcolor=\color{codegray},
    breaklines=true,
    showstringspaces=false,
    tabsize=4,
    aboveskip=1.0em,
    belowskip=1.0em,
    xleftmargin=0.5em,
    xrightmargin=0.5em,
    frame=single,
    framesep=0.5em    
}

\lstdefinestyle{myjava}{
    language=Java,
    basicstyle=\ttfamily\footnotesize,
    keywordstyle=\color{blue},
    commentstyle=\color{gray},
    stringstyle=\color{red},
    backgroundcolor=\color{codegray},
    breaklines=true,
    showstringspaces=false,
    tabsize=4,
    aboveskip=1.0em,
    belowskip=1.0em,
    xleftmargin=0.5em,
    xrightmargin=0.5em,
    frame=single,
    framesep=0.5em    
}

\lstdefinelanguage{JavaScript}{
    keywords={await, break, case, catch, class, const, continue, debugger, default, delete, do, else, export, extends, finally, for, function, if, import, in, instanceof, let, new, return, super, switch, this, throw, try, typeof, var, void, while, with, yield},
    keywordstyle=\color{blue},
    ndkeywords={boolean, null, true, false},
    ndkeywordstyle=\color{blue},
    identifierstyle=\color{black},
    sensitive=true,
    comment=[l]{//},
    morecomment=[s]{/*}{*/},
    commentstyle=\color{gray},
    stringstyle=\color{red},
    morestring=[b]',
    morestring=[b]"
}

\lstdefinestyle{myjavascript}{
    language=JavaScript,
    basicstyle=\ttfamily\footnotesize,
    keywordstyle=\color{blue},
    commentstyle=\color{gray},
    stringstyle=\color{red},
    backgroundcolor=\color{codegray},
    breaklines=true,
    showstringspaces=false,
    tabsize=4,
    aboveskip=1.0em,
    belowskip=1.0em,
    xleftmargin=0.5em,
    xrightmargin=0.5em,
    frame=single,
    framesep=0.5em    
}

\begin{document}

\title{Building a robust OAuth token based API Security: A High level Overview}

\author{Senthilkumar Gopal}

\maketitle

\begin{abstract}
APIs (Application Programming Interfaces) or Web Services are the foundational building blocks that enable interconnected systems. However this proliferation of APIs has also introduced security challenges that require systematic and scalable solutions for secure authentication and authorization. This paper presents the fundamentals necessary for building a such a token-based API security system. It discusses the components necessary, the integration of OAuth 2.0, extensibility of the token architectures, necessary cryptographic foundations, and persistence strategies to ensure secure and resilient operations. In addition to architectural concerns, the paper explores best practices for token lifecycle management, scope definition, expiration policies, and revocation mechanisms, all framed within a real-world scenario. By adhering to these principles, developers can establish a robust  baseline while maintaining the flexibility to customize their domain-specific requirements.

The approach does not claim to cover all variations necessary for diverse architectures but instead focuses on key principles essential for any standard API token authentication system. Throughout, the paper emphasizes balancing practical considerations with security imperatives and uses key concepts such as the CIA triad, OAuth standards, secure token life cycle, and practices for protecting sensitive user and application data. The intent is to equip developers with the foundational knowledge necessary to build secure, scalable token-based API security systems ready to handle the evolving threat landscape.
\end{abstract}


\section{Introduction}
\IEEEPARstart{A}{s} APIs increasingly serve as the connective tissue of modern software ecosystems, ensuring their protection is essential to maintain system integrity and user privacy. Token-based authentication has emerged as a widely adopted and effective method of securing APIs, with incidents such as the Panera Bread customer data leak \cite{krebs2018panerabread} and the Facebook OAuth access tokens \cite{meta2018securityupdate} serving as reminders of the outcomes when there are inadequate measures.

This paper presents a structured guide to building secure token-based API systems anchored in the core principles of security known as the CIA Triad - Confidentiality, Integrity and Availability. \textbf{Confidentiality} ensures that sensitive information is accessible only to authorized parties, \textbf{Integrity} guarantees that data remain unaltered and trustworthy throughout its lifecycle, while \textbf{Availability} ensures that the API services remain accessible to legitimate users, even in the face of attacks. These three form the foundation of all secure system design and are essential when building any API authentication infrastructure.

Despite the growing centrality of APIs in application development, many developers lack practical experience in building secure identity and authentication systems. Although domain-driven application design is well understood, security often remains an underexplored dimension. This paper seeks to close this gap by clarifying the essential needs of token infrastructures and how tokens assert identity, control access, and maintain the CIA triad in an interconnected landscape.

Unlike many commercial solutions that treat token systems as proprietary black boxes, this paper offers an open, pragmatic blueprint. It outlines foundational steps to establish a secure token architecture, focusing on OAuth 2.0 principles, cryptographic techniques, token life cycles, and best practices for sensitive data protection. The intent is not to cover every possible customization needed for specific industries, but to provide a strong and extensible foundation that developers can build upon.

In short, this paper aims to educate developers on how to treat security as a first-class concern in their API designs. By embracing these principles, developers can build authentication systems that are not only effective but also resilient, scalable, and transparent, forming a launchpad for more advanced and customized security solutions.

\section{Related Work}
Several recent studies have contributed to the field of token-based API security. \cite{banati2018authentication} proposed a novel approach for dynamic token validation in microservice architectures, addressing the challenges of maintaining consistent security policies across distributed systems. \cite{miksa2023enhancing} highlights the necessity of balancing scalability with strict validation protocols in environments where services evolve independently.

Other contributions have examined the application of zero-trust models to API security, focusing on continuous authentication and fine-grained authorization across microservice interactions \cite{NIST-SP-800-207}. Research into anomaly detection within token-based systems has also advanced, with \cite{liu2021anomalydetection} exploring machine learning techniques to identify malicious token behaviors through behavioral pattern analysis.

These works collectively inform the need for robust, adaptable, and privacy-preserving token architectures. However, much of the literature focuses on specific subdomains or theoretical models. In contrast, this paper aims to provide a comprehensive, component-based review suitable for practitioners who build production-grade API security systems.

\subsection{OWASP API Security Guidelines}
Aligning token and API security practices with the Open Web Application Security Project (\texttt{OWASP}) guidelines strengthen their defenses against common attack vectors. \texttt{OWASP} emphasizes minimizing default permissions, issuing only the scopes necessary for specific operations, and enforcing strict boundary controls between different scopes and access levels.

Multi-dimensional authorization schemes, which combine factors such as user identity, device fingerprint, and the risk level of the session, provide layered defense against excess-privilege or escalated access. This approach closely aligns with modern extensions to attribute-based access control (\texttt{ABAC}) models, offering context-aware and dynamic access evaluations beyond static role mappings.

By integrating \texttt{OWASP} recommendations and academic insights into system design, platforms can proactively harden their API ecosystems while preserving operational flexibility and scalability.

While the \texttt{OWASP} API Security Top Ten and associated guidelines provide an essential foundation for identifying common vulnerabilities and high-level mitigation strategies, this paper advances the discussion by focusing on deeper practical implementations and operational realities. This work helps identify architectural needs that extend beyond the scope of existing guidelines. 

It presents a detailed exploration of \textbf{real-world operational architectures}, addressing distributed systems challenges such as consistency models, auditing infrastructure, and scalable revocation strategies. These aspects are critical for implementing secure token-based authentication in large-scale, enterprise-grade environments. The paper extends the discussion of access control by introducing \textbf{fine-grained authorization models} that go beyond static scope enforcement. Drawing inspiration from hierarchical \texttt{RBAC} and context-aware \texttt{ABAC} models, the work supports modern adaptive access control requirements, allowing systems to dynamically adjust permissions based on session context.

The paper attempts to cover the end-to-end \textbf{token management life cycle and persistence engineering}, including token issuance, refresh, rotation, and revocation. It examines hybrid models like \emph{Phantom Tokens} \cite{sandoval2024phantom} and designs that balance performance, confidentiality, and operational resilience across API ecosystems. It addresses real life challenges such as \textbf{cryptographic key rotation automation}, offering practical architectural patterns for disruption-less key rollover processes. These practices enable secure, scalable cryptographic operations without service disruptions, which are not deeply covered in standard best-practice documents.

Finally, this work explores \textbf{elastic scalability and adaptive rate limiting} strategies. It analyzes real-time, dynamic approaches to request rate management that distinguish between legitimate high-traffic clients and attack scenarios, thus extending beyond \texttt{OWASP}'s traditional static threshold recommendations and meeting the needs of cloud-native, large-scale systems.

\section{Unique Security Challenges of APIs}
Web applications interact with users through user interfaces while APIs run the risk of exposing backend systems directly to external consumers and automated clients. This architectural difference introduces unique security challenges that extend beyond usual web application concerns. APIs operate on fine-grained data models and intend to allow programmatic access to system functions, intending to serve a variety of clients including mobile apps, IoT devices, and third party developer integrations with differing trust levels.

Web applications and its inputs are constrained by a defined user interface, while APIs must anticipate a broad range of requests, usage patterns, and potential abuse vectors. Attackers can easily automate DoS (\textit{Denial of Service}) attacks against API endpoints, bypassing authorization checks, probing for vulnerabilities or exploiting business logic flaws. Also, improperly managed API access tokens become exposed to attack vectors by enabling privilege escalation without tripping any conventional alerts.

The stateless nature of APIs demand security mechanisms that are distributed and detail oriented than those used in traditional monolithic applications. Hence, building secure APIs is not merely an extension of web application security; it requires distinct principles, models, and lifecycle management practices specific to the unique characteristics of API-driven architectures.

Note: This paper focuses primarily on the security of APIs. Wider trends around web application security, including web session management, UI driven vulnerabilities, browser based protections and threats are significant but beyond the scope of this discussion.

\section{Embracing Standards: The Case for OAuth 2.0}
Token based authentication through OAuth 2.0 framework currently acts as the de facto standard for securing APIs. OAuth 2.0 introduces a delegation protocol that enables resource owners to grant third party applications limited access to protected resources without directly sharing credentials. This separation of access and resources is fundamental to building scalable and a secure API ecosystem.

The OAuth protocol avoids anti-patterns, such as password sharing by enabling applications to act on behalf of users through scoped, time-limited tokens. Applications never directly handle user credentials, thereby minimizing the attack surface and improving their resiliency against credential theft.

Authentication verifies identity \textit{("Who are you?")}, while authorization determines permitted actions \textit{("What are you allowed to do?")}. OAuth primarily focuses on authorization. However, without robust authentication at the point of token issuance, all downstream authorization, rate limiting, and access controls can be compromised.

\begin{figure}[!t]
  \centering
  \includegraphics[width=\columnwidth]{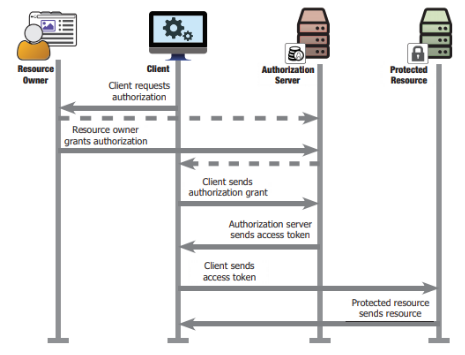}
  \caption{OAuth 2.0 authorization workflow. Adapted from~\cite{hardt2017oauth}.}
  \label{fig:oauth_flow}
\end{figure}

The OAuth 2.0 workflow involves several key actors and stages as indicated in Figure \ref{fig:oauth_flow}.

\noindent\textbf{Client Registration:} Applications register with the authorization server and receive a Client ID and Secret.

\noindent\textbf{User Authorization:} The client redirects the user to the authorization server for authentication and consent.

\noindent\textbf{Authorization Code Grant:} Upon consent, the server issues a temporary authorization code.

\noindent\textbf{Access Token Exchange:} The client exchanges the code for an access token via a secure server-to-server interaction.

\noindent\textbf{Access and Refresh Tokens:} Access tokens have short expiration periods; refresh tokens allow renewal without user re-authentication.

\noindent\textbf{Revocation Mechanisms:} OAuth supports token revocation to allow users to withdraw access at any time.

OAuth 2.1 draft \cite{hardt2023oauth21} attempts to further improve security guidelines, recommending short lived access tokens, mandatory Proof Key for Code Exchange (PKCE), and removal of the implicit grant type. OAuth formalizes boundaries between clients, resource servers, authorization servers, and resource owners, enforcing least privilege principles and enabling scalable, token based access control systems. This paper utilizes OAuth 2.0 as the foundation for broader token architectures and lifecycle strategies.

\subsection{Practical Integration: Language and Framework Support for OAuth Security}
Modern programming languages and frameworks simplify OAuth adoption by offering support for token management and security policies by default or with extensions.

\subsubsection{Java and Spring Security}

Spring Security provides OAuth 2.0 support for JVM languages. Developers can provide fine grained, method level authorization using annotations such as \texttt{@PreAuthorize} and \texttt{@PostAuthorize}. This approach provides both pre-execution and post-execution access control, enhancing authorization precision at the business logic layer. For instance:

\begin{lstlisting}[style=myjava, label={lst:spring_auth}]
@Service
public class AccountService {

    @PreAuthorize("@authz.decide()")
    @AuthorizeReturnObject
    public Account getAccountById(String accountId) {
        // Business logic here
    }
}
\end{lstlisting}
\noindent\textit{Fine-Grained Authorization with Spring Security}
\vspace{0.5em}

\subsubsection{Python and FastAPI}

FastAPI offers native OAuth 2.0 integration using dependency injection patterns. Developers can bind security scopes to API endpoints declaratively:

\begin{lstlisting}[style=mypython, label={lst:fastapi_oauth}]
from fastapi import FastAPI, Depends, Security
from fastapi.security import OAuth2PasswordBearer, SecurityScopes

app = FastAPI()
oauth2_scheme = OAuth2PasswordBearer(tokenUrl="token", scopes={"items:read": "Read items", "items:write": "Write items"})

@app.get("/items/")
async def read_items(security_scopes: SecurityScopes, token: str = Depends(oauth2_scheme)):
    # Your secure code here
\end{lstlisting}
\noindent\textit{OAuth2 Authentication Flow in FastAPI}
\vspace{0.5em}

This pattern provides fine grained access control along with token scopes with minimal overhead.

\subsubsection{JavaScript and Node.js/Express}
In JavaScript, frameworks such as Express integrate OAuth security via middleware libraries like \mbox{\texttt{express-oauth2-jwt-bearer}}. Middleware automatically validates tokens and enforces audience, issuer, and signing algorithm constraints and such approaches maintain centralized security enforcement across API routes, improving maintainability and reducing errors.

\begin{lstlisting}[style=myjavascript, label={lst:express_oauth_js}]
const { auth } = require('express-oauth2-jwt-bearer');
const express = require('express');
const app = express();

const checkJwt = auth({
  audience: 'https://my-api.com',
  issuerBaseURL: `https://YOUR_DOMAIN/`,
});

app.use(checkJwt);

app.get('/api/private', (req, res) => {
  res.send('Hello from a private endpoint!');
});
\end{lstlisting}
\noindent\textit{OAuth2 JWT Authentication Middleware in Express.js}

These are a few illustrative examples of programming language level support provided for ease of integration and utilization of token based OAuth security.

\section{Designing Secure Token Architectures}
As APIs rely on token based mechanisms for authentication and authorization, designing robust, extensible, and secure token structures are critical and a poorly designed token can expose sensitive information, introduce authorization bypasses, or compromise system integrity. In modern API security ecosystems, \textbf{JSON Web Tokens (JWTs)} have emerged as a leading standard for token formats due to their compactness, flexibility, and cryptographic security features.

\subsection{JSON Web Tokens (JWT): Foundation of Secure Token Design}

\textbf{JSON Web Tokens (JWTs)} have gained widespread adoption as a compact and self contained structure for securely transmitting information between parties as a JSON object. JWTs are particularly suited for OAuth 2.0 implementations, where they often serve as access tokens due to their ability to encapsulate claims and be cryptographically signed. Their structure allows security systems to verify a token’s integrity and trustworthiness without requiring persistent server-side sessions.

\begin{figure}[!t]
  \centering
  \includegraphics[width=\columnwidth]{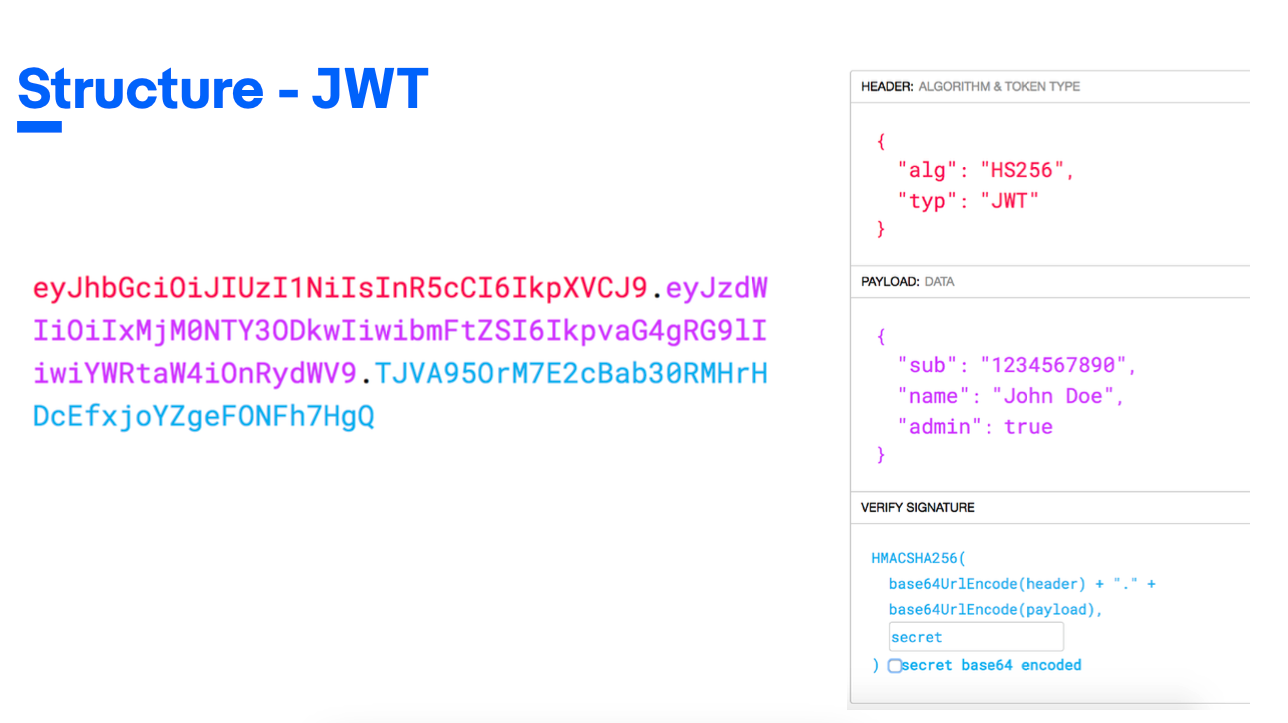}
  \caption{Structure of a JWT Token}
  \label{fig:jwt_structure}
\end{figure}

JWT typically consists of three distinct components as illustrated in Figure \ref{fig:jwt_structure}.

\noindent\textbf{Header:} with metadata about the token, including the type of token (typically "JWT") and the hashing algorithm used for signing (e.g., \texttt{RS256} for asymmetric keys or \texttt{HS256} for symmetric secrets).

\noindent\textbf{Payload (Claims):} contain the assertions and \textit{"claims"} about the entity, such as user identity, application identity, and access permissions. They may also contain optional metadata like token versioning.

\noindent\textbf{Signature:} provides cryptographic proof that the token was issued by a trusted authority and that its contents have not been tampered with.

Whike designing their JWT header, developers should follow best practices by explicitly specifying the token type (\texttt{typ}) and secure/vetted signing algorithms (\texttt{alg}). Use of "none" as an algorithm or insecure algorithms (e.g., \texttt{HS256} when asymmetric signing is expected) must be avoided, as these are common attack vectors.

\subsection{Example of a Secure JWT Payload}

A secure, extensible JWT payload might look like the following:

\begin{lstlisting}[style=myjavascript, label={lst:jwt_struct}]
{
  "sub": "1234567890",
  "aud": "api.example.com",
  "iss": "auth.example.com",
  "exp": 1712040000,
  "iat": 1712036400,
  "scope": "read:customers write:orders",
  "app_id": "ecommerce-app",
  "device_id": "device-8873abc",
  "ip": "203.0.113.42",
  "ver": "1.0"
}
\end{lstlisting}

This structure ensures the inclusion of essential claims such as user identity (\texttt{sub}), application association (\texttt{app\_id}), scope of authorization (\texttt{scope}), time-bounded validity (\texttt{exp}, \texttt{iat}), device metadata, and token versioning (\texttt{ver}). Embedding structured and minimal claims enables scalable authorization enforcement without leaking unnecessary sensitive information.

\subsection{Designing Claims and Metadata}

Developers should consider what information is embedded inside the JWT payload. Key categories of claims include:

\noindent\textbf{User Identity:} uniquely identifies the authenticated user.

\noindent\textbf{Application Identity:} links the issued token to the requesting client application.

\noindent\textbf{Authorization Scopes:} specify the level of access permitted.

\noindent\textbf{Issuance and Expiration Times:} control token validity windows.

\noindent\textbf{Device and IP Information:} help detect session hijacking and misuse patterns.

\noindent\textbf{Token Versioning:} facilitates schema upgrades and backward compatibility.

Recent research \cite{alaca2016device} indicate that including dynamic context information, such as device fingerprints or login locations can significantly enhance token resilience without materially impacting performance. Its essential to maintain an extensible token structure to ensure future system upgrades and threat model changes should be absorbable without invalidating existing issued tokens.

\subsection{Confidentiality and the Role of Claims}

Though \texttt{JWT}s are \textbf{signed} to ensure \textbf{integrity}, their contents are \texttt{base64url}-encoded and not encrypted. So anyone in possession of a \texttt{JWT} can decode and inspect the payload, unless additional encryption is applied. This requires thoughtful claim design and sensitive data (such as personally identifiable information or internal authorization flags) should not be exposed. Encryption should be considered for usecases where claim confidentiality is needed. Following \textbf{CIA triad} principles of \textbf{Confidentiality}, \textbf{Integrity}, and \textbf{Availability} is essential while balancing data to include within tokens along with data that should be protected at transport or storage layers.

\subsection{Note on JWE vs JWT}

There is a critical distinction between signed tokens (\texttt{JSON Web Signature}, \texttt{JWS}) and encrypted tokens (\texttt{JSON Web Encryption}, \texttt{JWE}). \texttt{JWS} provides a mechanism where the token is signed using a cryptographic algorithm to ensure its integrity and authenticity. While the contents of a \texttt{JWS} can be read by anyone who has possession of the token, the signature allows the recipients to verify that the token has not been tampered with and is genuinely issued by a trusted issuer. \texttt{JWE} extends this beyond ensuring integrity by encrypting the token’s contents, which safeguards against unauthorized access. This encryption ensures that the contents of the token can only be read by the intended recipients who possess the necessary decryption keys.

\noindent\textbf{Use JWS (Signed Tokens):} when there is no confidentiality requirement of the information in the token and the primary objective is to verify authenticity. This is the most common scenario where the token's payload does not contain sensitive information and can be used in environments where interception of tokens is not a security concern.

\noindent\textbf{Use JWE (Encrypted Tokens):} when the token contains sensitive data that should not be exposed to unauthorized parties. Encryption is essential in high-risk environments or applications where tokens carry personally identifiable information (PII), financial details, or other confidential data. This is also essential when tokens are transmitted over channels that are susceptible to interception.

The choice between \texttt{JWS} and \texttt{JWE} is determined by the sensitivity of the payload’s content and the security requirements of the application environment.

\subsection{Cryptographic Considerations}

Selecting the appropriate \textbf{signing algorithm} is important to protect the integrity of tokens. HMAC based algorithms are effective when the issuer and verifier of the token are the same entity, as the security of the token depends on the secrecy of the shared key. For systems where the issuer and verifier are distinct, asymmetric algorithms such as \texttt{RSA} or \texttt{ECDSA} are preferable due to their use of separate keys for signing and verification, enhancing security in distributed environments.

Following best practices for key management are critical for the continued security of the token system, which include the scheduled rotation of keys to mitigate the risk of compromise, secure storage mechanisms to prevent unauthorized access, and careful distribution of public keys. For example, Amazon Web Services (AWS) implements advanced key management protocols through AWS Key Management Service (\texttt{KMS}). \texttt{AWS KMS} allows users to create and control the encryption keys used to encrypt their data, providing secure key storage, key rotation, and auditing capabilities to ensure that the keys are handled securely throughout their lifecycle.

\subsection{Privacy-Preserving Token Structures and Future Directions}

While token based authentication systems such as \texttt{OAuth 2.0} prioritize integrity and availability, maintaining confidentiality of token claims is also becoming increasingly important, especially in the context of data protection regulations. As discussed earlier,  standard \texttt{JWT}s, are signed and not encrypted allowing their contents visible to anyone in possession of the token and needs attention where sensitive user information, authorization flags, or resource scopes are embedded within token payloads.

Recent advancements have explored the integration of cryptographic techniques to address this limitation. \cite{cherif2023zero} introduced a framework for \textbf{confidential and auditable OAuth tokens} using zero-knowledge proofs (\texttt{ZKP}s). This system allows resource servers to verify specific claims about a token (e.g., permission to access a resource) without revealing the full contents of the token itself. By embedding verifiable cryptographic proofs inside token structures, \texttt{ZKP}-based models achieve both confidentiality and ability to audit without sacrificing performance.

Integrating such techniques into token architecture enable platforms to minimize data exposure even in distributed systems where tokens traverse multiple intermediaries. It also enhances compliance with data privacy regulations such as \texttt{GDPR} and \texttt{CCPA}, which increasingly mandate data minimization and propose limitation principles.

Future token systems should consider augmenting traditional by-value or by-reference models with privacy preserving designs. Though there might be operational and implementation complexity, having a combined system for verifiability, confidentiality, and scalability marks a significant advancement in secure token-based API design. Further research could explore how zero-knowledge  tokens interact with scalable revocation models and adaptive access control mechanisms, ensuring that confidentiality enhancements do not inadvertently introduce new attack surfaces.

\section{Token Life Cycle Management}
\begin{figure}[!t]
  \centering
  \includegraphics[width=\columnwidth]{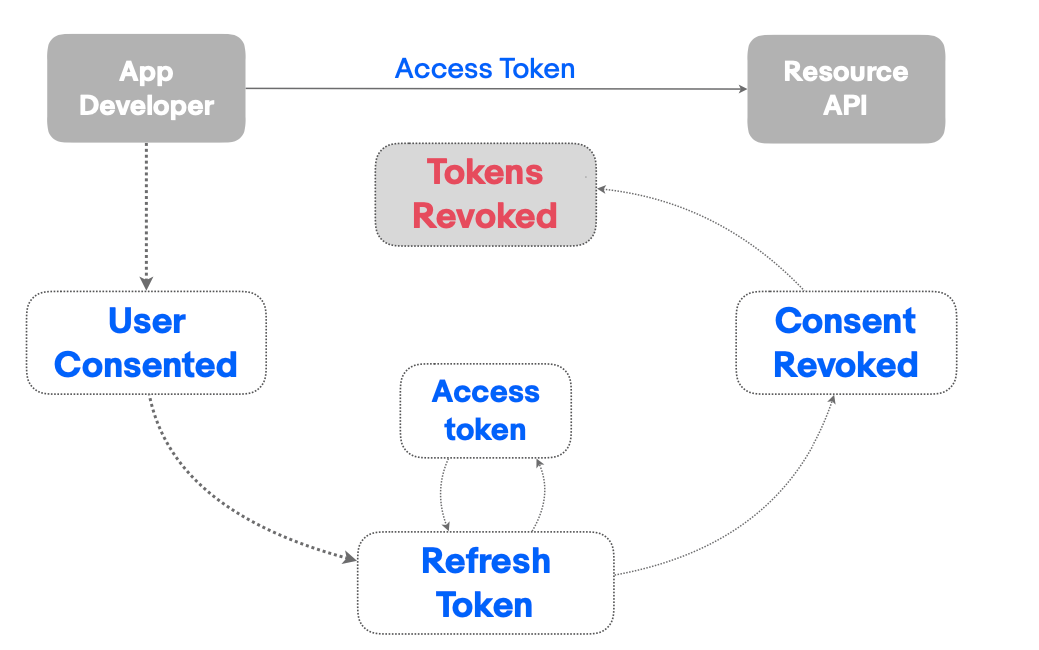}
  \caption{A simple representation of a token lifecycle}
  \label{fig:token_life_cycle}
\end{figure}

API security depends on effectively managing the life cycles of users, applications, and tokens. The life cycle management of a token involves several key phases as depicted in Figure \ref{fig:token_life_cycle}.

\noindent\textbf{User Registration and Activity:} stage ensures that user authentication processes are secure and efficient. Proper validation during user registration is crucial as it sets the groundwork for subsequent security measures. For a deeper discussion on user life cycle management in digital systems, refer to \cite{thakur2015user}, who explores user identity verification techniques in various technological environments.

\noindent\textbf{Token Expiration and Revocation:} is vital to manage the validity period of tokens to mitigate risks associated with token compromise. Tokens should have a defined lifespan after which they expire and require renewal. Effective token revocation mechanisms are essential for invalidating compromised tokens quickly, in order to response to security breaches.

\noindent\textbf{Token Rotation:} for regularly updating cryptographic keys and tokens is a security best practice that helps safeguard against potential vulnerabilities. This process, known as token rotation, involves issuing new tokens at predefined intervals or under specific conditions, thus maintaining the security integrity over time. \cite{ferry2015security} describe methods for integrating secure and automated token rotation systems within enterprise applications.

\subsection{Application Lifecycle for a Platform Provider}
\begin{figure}[!t]
  \centering
  \includegraphics[width=\columnwidth]{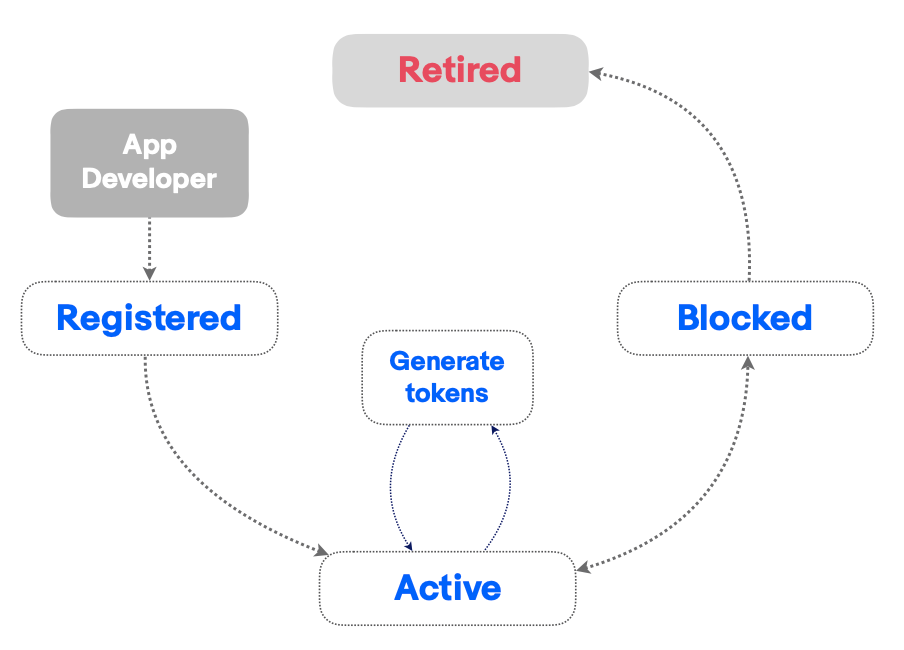}
  \caption{Lifecycle stages of an application}
  \label{fig:app_lifecycle}
\end{figure}

A registered application goes through multiple phases as depicted in Figure \ref{fig:app_lifecycle}. As a platform provider offering \texttt{OAuth} capabilities, managing the application lifecycle registered by developers is critical for maintaining strong API security. This lifecycle management involves several key stages to ensure comprehensive security such as:

\noindent\textbf{Developer Registration and Application Setup:} Security begins when developers register and set up their \texttt{OAuth} applications on the platform. The platform must enforce strict security standards during registration - requiring strong authentication and secure communication protocols. This ensures that only verified developers can register applications, thereby safeguarding the platform from malicious entities.

\noindent\textbf{Security Reviews and Approval:} Applications should undergo thorough security reviews before being granted access to \texttt{OAuth} capabilities. This includes reviewing the application’s architecture, security features, and compliance with data protection laws. Applications meeting the platform's security criteria are approved for deployment and such review processes help prevent potential vulnerabilities that could be exploited once the application is live.

\noindent\textbf{Token Generation and Management:} As an \texttt{OAuth} provider, the platform handles the generation and management of access and refresh tokens and should implement secure token generation practices, enforce token expiration policies, and provide mechanisms for token revocation. This control over the token lifecycle ensures that tokens are used appropriately and reduce the risk of unauthorized access.

\noindent\textbf{Monitoring and Compliance:} Continuous monitoring of application interaction with platform’s APIs is crucial. This includes tracking unusual activity patterns and ensuring applications adhere to the platform's usage policies. Regular compliance audits help maintain high security standards and enforce policy adherence, protecting both the platform and its users.

\noindent\textbf{Decommissioning and Offboarding:} When an application is no longer needed or if a developer wishes to discontinue their service, a secure decommissioning process is important. This involves revoking all active tokens, securely deleting sensitive data, and confirming that no back doors are left open. A structured offboarding process prevents data leaks and ensures that the application’s deprecation is handled securely.

This proactive approach to application lifecycle management from a platform provider's perspective ensures that all applications using \texttt{OAuth} capabilities adhere to stringent security standards, thereby minimizing risks across the ecosystem.

\subsection{Issuance and Validation}

Tokens must be issued taking application's security state into account and the specific permissions granted to a user. Each API request submitted along with the token should verify the token’s expiration, signature authenticity, and the appropriateness of included claims. This ensures that only valid, authorized requests are processed, thereby enhancing the security of the API ecosystem. The principles of secure token issuance and validation are elaborated in the work of \cite{siriwardena2014advanced}, who examine the security frameworks for API interactions.

To optimize security while maintaining user sessions, a well structured \textbf{Refresh Mechanism} for token renewal is required. This system allows for the use of shorter lived access tokens, which reduces the window of opportunity for attackers in case a token is compromised. Refresh tokens support the regeneration of access tokens, thereby extending user sessions securely without the need for frequent re-authentications. This approach is discussed in \cite{ethelbert2017json}, who analyze refresh token strategies to enhance session security in cloud services.

A robust \textbf{token revocation} system is important for maintaining the integrity of the API security architecture. Effective revocation strategies may involve maintaining a blocklist of revoked tokens or adopting a stateful token approach, where each token’s validity is continually verified against a central database or authority. Implementing dynamic revocation methods helps manage security incidents effectively, as explored by \cite{ferry2015security} in their study on stateful versus stateless token revocation systems in distributed networks.

\section{Persistence and Scalability}
The next crucial aspect is the persistence strategy for token management and the data architecture of the authorization server. The design considerations for persistence fundamentally impact the operational integrity and efficiency of token-based systems. An effective authorization server architecture must incorporate mechanisms for caching, robust databases, and efficient metadata management. These are pivotal for maintaining high availability and security of the token services. Ensuring \textbf{atomic consistency} is crucial for operations such as token revocation and claim updates. This is necessary to maintain the trust and integrity of the authentication system, where immediate consistency is required to reflect changes across all nodes without delay.

\subsection{By-Value vs. By-Reference Tokens}

\begin{figure}[!t]
    \centering
    \includegraphics[width=\columnwidth]{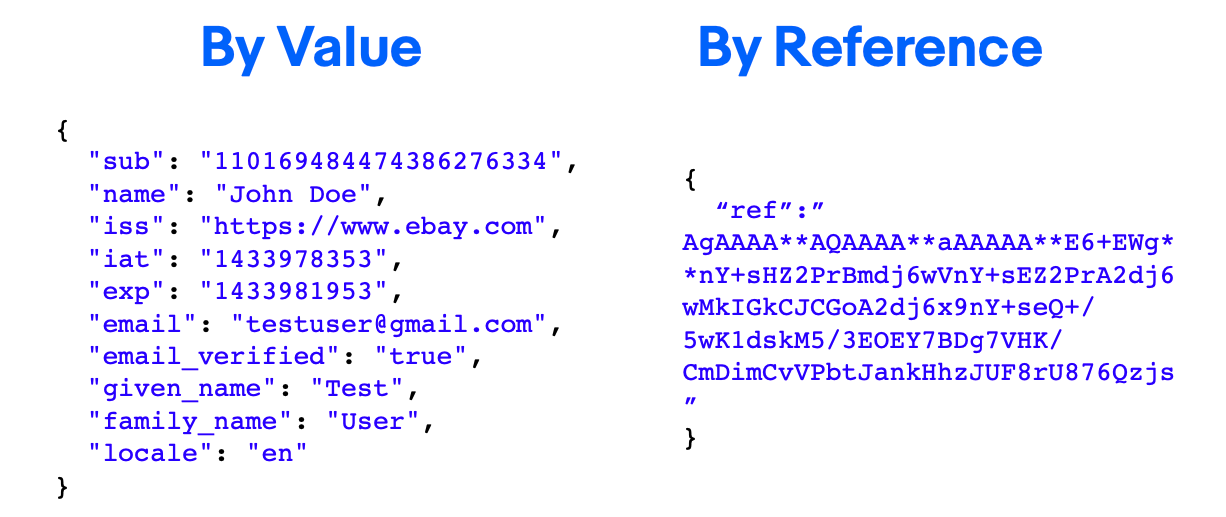}
    \caption{Comparison of By-Value and By-Reference Token Architectures}
    \label{fig:ref_value_comp}
\end{figure}

\noindent \textbf{By-Value Tokens} like \texttt{JWT}s that carry all required data allow for stateless operations by resource servers, which can process authentication without additional database lookups. This enhances performance due to reduced latency and improves scalability by minimizing reliance on server-side state management. However, the challenge arises when the payload of these tokens becomes large, potentially impacting transmission times and increasing storage requirements. Balancing the payload content is essential to maintain performance efficiency.

\noindent \textbf{By-Reference Tokens} Alternatively, by-reference tokens serve as pseudo identifiers that require resource servers to perform database lookups to fetch the associated data for each token presented. This model increases latency due to additional network or IO calls but offers better control over token revocation and enhances confidentiality by storing sensitive data server-side rather than within each token.

An illustrative example of By-Value and By-Reference tokens is provided in Figure \ref{fig:ref_value_comp}

\subsection{Hybrid Token Management Approaches}

Considering the tradeoffs between By-Value and By-Reference tokens, a hybrid approach often represents a balanced solution. For example, the \emph{Phantom Token} \cite{sandoval2024phantom} approach leverages the benefits of both models by issuing opaque tokens to clients and converting them to \texttt{JWT}s at an API gateway. This strategy optimizes security and operational efficiency by utilizing server-side token introspection to validate and translate tokens. While by-value tokens provide significant performance benefits due to their independence from server-side state, incorporating some stateful elements through server-side validation can enhance security without substantially compromising performance.

\subsection{Database Considerations for Stateful Tokens}

Choosing the appropriate database is critical for managing stateful aspects of the token system. Important considerations include read/write performance, scalability, and consistency requirements. Recent advancements, such as those discussed by \cite{rodriguez2001analysis}, highlight the use of distributed caching systems to enhance the performance and scalability of token storage in high-traffic API environments.

Balancing token size and payload content in by-value token systems, such as \texttt{JWT}s, requires careful design to ensure both performance and sufficient information for authentication. The payload should include only essential information, such as a unique user identifier, token expiration time, and minimal role or scope data necessary for authorization decisions. Large datasets, detailed user profiles, or session-specific metadata should be excluded from the token and retrieved dynamically if needed. 

Additionally, applying data compression techniques can reduce token size during transmission, although the computational cost of compression and decompression must be evaluated against network performance benefits. Structuring claims efficiently, using short standardized field names, and avoiding redundancy further helps minimize token size. Where possible, detailed access rights should be abstracted into broader scopes or permission levels to reduce payload complexity while maintaining flexible authorization control. By applying these practices, developers can maintain lean tokens that perform well without sacrificing the necessary security context required for authentication and access control.

\subsection{Security Considerations for By-Reference Tokens}

By-Reference tokens introduce additional security risks that need to be managed. By-Reference tokens act as opaque identifiers requiring resource servers to query the authorization server for user or session data and introduces additional latency and load on the backend systems. This creates a strong dependency on the availability and performance of the authorization server and marking the centralized store that holds the token metadata as a critical security target which could \textit{potentially} expose sensitive user information even if the tokens themselves remain opaque.

To mitigate these risks, caching token metadata securely at the resource server can help reduce the frequency of remote lookups and cache invalidation strategies must be delicately handled to ensure token revocations and updates are working in a consistent manner. Securing all communications between resource servers and authorization servers with strong protocols such as \texttt{TLS} is mandatory to prevent interception. Finally, the authorization server and its data store must be hardened through strict access controls, auditing, and active intrusion monitoring as well.

Designing the authentication infrastructure to be horizontally scalable and resilient to load spikes is critical to ensure that increased token introspection does not become a bottleneck or single point of failure. Effective persistence strategies requires a careful balance between performance, security, and scalability. Each choice of adopting by-value, by-reference, or a hybrid token approach must be correctly managed to support the overarching goals of the API security framework.

\section{Scopes and Fine-Grained Authorization Control}

Design and enforcement of scopes identify and define the specific actions or resources a token holder is authorized to access. Implementing fine-grained scopes enables precise control over access rights, significantly reducing the potential attack surface in case a token is compromised. Each scope should correspond to a narrowly defined permission, avoiding broad or ambiguous grants of authority.

The design of scopes is closely related to the principles of Role-Based Access Control (RBAC), a widely adopted authorization model. \texttt{RBAC} assigns permissions to roles rather than individual users. Users are then associated with one or more roles based on their responsibilities within an organization. This abstraction simplifies administration and ensures that permissions are granted consistently and according to the principle of least privilege. \cite{ferraiolo1995role} first formalized \texttt{RBAC} as a flexible and scalable model for managing large-scale access control systems, highlighting its effectiveness in minimizing the risk of over-privileged accounts.

Effective scope design often benefits from hierarchical organization, where broader scopes can inherit permissions from more specific scopes. Hierarchical \texttt{RBAC}, an extension described \cite{bertino2003rbac}, introduces role hierarchies to allow senior roles to inherit permissions from junior roles, thereby supporting scalable permission management. Applying similar principles to scopes within API systems allows for modular permission structures that accommodate diverse access needs while preserving strict control.

Scopes must be actively maintained to remain effective. Regular auditing of scope definitions ensures that access permissions are current and aligned with evolving business and security requirements. Outdated or overly permissive scopes alongside their access to resources should be revised or deprecated in a periodic manner to prevent unnecessary access vectors. Studies such as \cite{sandhu1994access} further emphasize that regular review of access control assignments is essential to maintaining a secure operational environment.

In addition to careful scope design, minimizing both the breadth of scopes and the lifespan of tokens is fundamental to reducing exposure. Tokens should be issued with the minimal set of scopes necessary for the task, avoiding defaults that grant excessive privileges. Token expiration should also be aggressively managed as tokens with minimal lifespan reduce the window of opportunity for misuse if compromised. Longer lived credentials, such as refresh tokens, must be subject to stricter validation policies and monitoring and should be structured for minimal usage across network to prevent loss via MITM \textit{(Man In the Middle Attacks)}. Following the principles of minimal privilege and constrained lifespan, as outlined in the broader \texttt{RBAC} literature, strengthens the resilience of \texttt{OAuth}-based systems against unauthorized access.

\section{System Integrity and Scale Considerations}

\subsection{Token Revocation and Recovery Strategies}

After the implementation of fine-grained scopes and minimal authorization, the ability to revoke tokens efficiently in response to security incidents is necessary which can support revocation at multiple levels to address different risk scenarios.

\noindent\textbf{User-level token revocation:} targets specific user credentials that may have been compromised without affecting other users.

\noindent\textbf{Application-level token revocation:} addresses cases where an entire application must be deauthorized due to compromise or policy violations.

\noindent\textbf{System-wide emergency revocation:} enables the platform to globally invalidate all active tokens in response to critical incidents, such as breaches or cryptographic failures.

Designing revocation strategies to minimize operational disruption during system wide emergency responses requires careful planning of storage and consistency models. Strong transactional guarantees, consistent revocation propagation, and fail-safe defaults are critical to ensure unverified tokens are treated as invalid. Facebook’s revocation of over 90 million user tokens \cite{meta2018securityupdate} following a breach demonstrates the importance of scalable revocation infrastructure backed by robust consistency guarantees.

Work by \cite{sun2012systematically} systematically analyzed practical limitations of existing \texttt{OAuth 2.0} and \texttt{OpenID Connect} revocation mechanisms. This study highlighted that standard token revocation endpoints often suffer from scalability issues and latency under load. They propose enhanced models based on scalable revocation lists and distributed token status services, where resource servers periodically sync lightweight revocation proofs instead of querying introspection endpoints for each request. Integrating such mechanisms into the revocation architecture allows systems to minimize reliance on synchronous validation while preserving strong revocation guarantees.

Keeping the systems up to date with such techniques can improve revocation responsiveness where short-lived tokens are combined with revocation proof synchronization. These designs maintain near realtime revocation status while significantly reducing backend bottlenecks during mass revocation events. In cloud native environments, such improvements are critical to achieving emergency revocation at scale without disrupting normal API traffic.

\section{Cryptographic Key Management and Recovery}

Token revocation by itslef is insufficient without corresponding key management and rotation practices. Regular cryptographic key rotation is critical to limit the risk window associated with long-term key exposure. Platforms must automate key rotation workflows, ensuring that new keys are generated, distributed, and propagated with minimal human intervention. Automated key management frameworks such as \cite{das2012key} illustrate the necessity of seamless rotation at scale.

Architectural designs that support automated key rotation without service disruption include publishing new public keys ahead of activation, supporting key rollover periods where old and new keys are both accepted, and embedding version identifiers (\texttt{kid} fields) inside tokens. These patterns enable smooth transition during key rotations and guarantee that services can continue validating tokens both pre and post rotation without interruptions or loss of authentication fidelity.

\section{Auditing, Monitoring, and Fingerprinting for Anomaly Detection}

Continuous visibility into token activity is another fundamental component for detecting and responding to security threats. Comprehensive auditing must have capability to capture every token issuance, usage, refresh, and revocation events to construct a complete lifecyle and access history. Monitoring access patterns enables early identification of abnormal behaviors, such as anomalous API call volumes, unexpected geographic access, or device fingerprint mismatches.

Architectural trade-offs exist between centralized and distributed audit models. Centralized systems provide a unified and consistent view of access activities, simplifying correlation and policy enforcement. However, they can become performance bottlenecks and introduce single points of failure under high traffic loads. Distributed monitoring systems enhance resilience and scalability by aggregating events closer to the source. Techniques such as those outlined by \cite{benson2020disco} enable efficient distributed event aggregation, maintaining comprehensive auditing without sacrificing performance.

Recent research has introduced machine learning-based anomaly detection into token monitoring systems. \cite{luo2021deep} proposed sequential models, including \texttt{LSTM}s and RNN based architecture, to analyze streams of \texttt{OAuth} interactions. Unlike static rule based systems, these models learn normal behavioral sequences of token issuance, refresh, and usage, enabling the detection of deviations that may indicate token theft, misuse, or session hijacking. Modeling behavioral sequences rather than isolated events improves detection rates and reduces false positives compared to traditional heuristics. A detailed deep dive into this topic is beyond the scope of this paper, however building robust monitoring systems for token utilization is a key component for maintaining the security and continued availability of API security.

Additional advancements incorporate behavioral analytics and heuristic based detection. Behavioral analytics monitor typical API call frequencies and timings, flagging deviations for further inspection \cite{ranjan2022user}. Heuristic based detection relies on predefined rules to quickly identify signs of misuse, such as repeated failed authentication attempts. Signature based detection, which identifies known attack vectors, complements these methods, requiring regular updates to remain effective.

Tools like \texttt{OAuthTester} simulate attack scenarios to detect vulnerabilities in \texttt{OAuth 2.0} implementations using adaptive model based security testing \cite{yang2016model}. Static analysis tools such as \texttt{Cerberus}, developed by \cite{rahat2022cerberus}, identify logical flaws in \texttt{OAuth} service provider libraries, emphasizing the role of heuristic based techniques in uncovering implementation vulnerabilities.

Integrating machine learning driven anomaly detection, distributed event aggregation, fingerprint binding, and heuristic based methods forms a resilient monitoring architecture. Techniques such as fingerprinting tokens to device identifiers, browser fingerprints, or IP ranges further strengthen detection capabilities. Tokens used outside their expected context can be flagged and revoked preemptively, minimizing the window of exposure.

Together, these approaches enable platforms to shift from reactive to proactive defense and detect credential misuse at scale, without overwhelming the operation teams. Recent contributions by \cite{ramamoorthi2024anomaly} highlight the growing importance of machine learning for anomaly detection in API security, underscoring the need for continuous innovation to protect \texttt{OAuth} based systems.

\section{Scalability, Rate Limiting, and Resilience Engineering}

As platform usage grows, API security mechanisms must scale proportionally to maintain operational integrity under increasing load. Designing for scalability requires effective rate limiting to control request volumes and defend against abuse patterns, particularly distributed denial-of-service (DDoS) attacks at the application layer.

Traditional rate limiting relies on static thresholds, which can be either too restrictive during normal surges in traffic or too permissive during targeted attacks. Recent research by \cite{serbout2023api} proposes elastic rate limiting strategies that dynamically adjust rate limits based on real time application metrics, historical client behavior, and risk scoring models. Unlike static thresholds, elastic models allow systems to intelligently differentiate between legitimate high-volume clients and malicious traffic patterns.

Fine tuned adaptive rate limiting to balance legitimate high traffic use cases against DDoS resilience involves establishing multiple tiers of trust. Verified applications or long term trusted clients may be granted higher dynamic thresholds, while unknown or newly registered clients operate under stricter baseline limits. Systems can incorporate feedback loops that monitor success rates, error patterns, and latency anomalies to continuously reclassify clients and adapt limits accordingly. Integrating predictive analytics, as proposed by \cite{mathur2024artificial} further strengthens defenses by anticipating traffic anomalies before they escalate into service degradation.

Alongside rate limiting, maintaining atomic consistency during critical operations, such as token revocation and permission updates, is necessary for security correctness in distributed systems. Brewer’s CAP theorem \cite{simon2000brewer} highlights the need to prioritize consistency over availability in security-sensitive operations to avoid race conditions or stale authorization states.

Resilient architectures must also automate cryptographic operations such as key rotation and token introspection updates, ensuring security processes scale automatically with system demand. Together, elastic rate limiting, consistency aware database operations, and automated cryptographic practices form the foundation of a scalable and resilient API security infrastructure.

\section{Conclusion}
This paper has presented a comprehensive framework for securing token based API systems, grounded in established protocols, practical architectural principles, and proactive operational strategies. \textbf{Robust API security} demands adherence to best practices and continuous evolution in response to emerging threats and system requirements.

We emphasized that secure design begins with adopting standard protocols such as \texttt{OAuth 2.0} and \texttt{OpenID Connect}, avoiding reliance on security through obscurity. Critical foundations include designing \textbf{extensible, structured token models}, implementing \textbf{minimal and well defined scopes}, and enforcing \textbf{strong authentication and revocation mechanisms}. Managing the token lifecycle, automating key rotation, reducing token exposure, and ensuring atomic consistency under high load form the operational pillars of resilient API platforms.

Recent advances in the field further enhance this blueprint. \textbf{Scalable revocation mechanisms} based on distributed revocation proofs provide a path to maintain token security even during mass revocation events. \textbf{Machine learning driven anomaly detection models}, capable of analyzing sequential \texttt{OAuth} usage patterns, enable early detection of even sophisticated credential misuse. \textbf{Elastic rate limiting strategies} dynamically balance the needs of legitimate high traffic clients against the resilience requirements of API services under potential DDoS attacks. Innovations in \textbf{privacy preserving token structures}, such as the application of zero knowledge proofs, allow claims to be verified without exposing sensitive token contents, addressing critical data confidentiality and compliance challenges.

\textbf{API security}, like broader cybersecurity, remains an evolving contest between defenders and adversaries. The \textbf{ten principles} outlined in this work, ranging from secure token architecture to continuous auditing and automation, offer a practical and scalable foundation for securing modern API ecosystems. Platforms must continually audit, refine, and adapt their systems to meet new threats, regulatory shifts, and evolving architectural demands.

Future work should focus on operationalizing \textbf{machine learning based anomaly detection at scale}, developing \textbf{standardized evaluation metrics for token security performance}, and integrating \textbf{privacy preserving authentication techniques} into mainstream \texttt{OAuth} implementations. Further research into scalable, decentralized revocation and fine grained, context-aware authorization models will be necessary to secure API infrastructures against increasingly sophisticated and distributed attack vectors.

\section*{Acknowledgments}

The author acknowledges the use of OpenAI's ChatGPT (Mar 2024 version) in the preparation of this manuscript. \cite{chatgpt} was utilized to assist in correcting grammar, enhancing clarity, and rewriting sections to improve readability. All technical content, analysis, and conclusions presented in this paper were solely developed by the author. 

The use of artificial intelligence (AI)–generated assistance is disclosed in accordance with IEEE policies regarding the use of AI tools in scholarly publications.

\bibliographystyle{IEEEtran}
\bibliography{token_api_security_ieee}

\begin{thebibliography}{10}
\providecommand{\url}[1]{#1}
\csname url@samestyle\endcsname
\providecommand{\newblock}{\relax}
\providecommand{\bibinfo}[2]{#2}
\providecommand{\BIBentrySTDinterwordspacing}{\spaceskip=0pt\relax}
\providecommand{\BIBentryALTinterwordstretchfactor}{4}
\providecommand{\BIBentryALTinterwordspacing}{\spaceskip=\fontdimen2\font plus
\BIBentryALTinterwordstretchfactor\fontdimen3\font minus \fontdimen4\font\relax}
\providecommand{\BIBforeignlanguage}[2]{{%
\expandafter\ifx\csname l@#1\endcsname\relax
\typeout{** WARNING: IEEEtran.bst: No hyphenation pattern has been}%
\typeout{** loaded for the language `#1'. Using the pattern for}%
\typeout{** the default language instead.}%
\else
\language=\csname l@#1\endcsname
\fi
#2}}
\providecommand{\BIBdecl}{\relax}
\BIBdecl

\bibitem{krebs2018panerabread}
\BIBentryALTinterwordspacing
B.~Krebs, ``Panerabread.com leaks millions of customer records,'' April 2018. [Online]. Available: \url{https://krebsonsecurity.com/2018/04/panerabread-com-leaks-millions-of-customer-records/}
\BIBentrySTDinterwordspacing

\bibitem{meta2018securityupdate}
\BIBentryALTinterwordspacing
{Meta}, ``{Security Update},'' September 2018. [Online]. Available: \url{https://about.fb.com/news/2018/09/security-update/}
\BIBentrySTDinterwordspacing

\bibitem{banati2018authentication}
A.~B{\'a}n{\'a}ti, E.~Kail, K.~Kar{\'o}czkai, and M.~Kozlovszky, ``Authentication and authorization orchestrator for microservice-based software architectures,'' in \emph{2018 41st International Convention on Information and Communication Technology, Electronics and Microelectronics (MIPRO)}.\hskip 1em plus 0.5em minus 0.4em\relax IEEE, 2018, pp. 1180--1184.

\bibitem{miksa2023enhancing}
T.~Miksa and Others, ``Enhancing microservices security with token-based access control,'' \emph{Journal of Cloud Computing}, vol.~12, no.~1, pp. 1--15, 2023.

\bibitem{NIST-SP-800-207}
\BIBentryALTinterwordspacing
S.~Rose, O.~Borchert, S.~Mitchell, and S.~Connelly, ``Zero trust architecture,'' National Institute of Standards and Technology, Special Publication 800-207, August 2020. [Online]. Available: \url{https://nvlpubs.nist.gov/nistpubs/SpecialPublications/NIST.SP.800-207.pdf}
\BIBentrySTDinterwordspacing

\bibitem{liu2021anomalydetection}
H.~Liu and Others, ``Enhancing api security through machine learning-based anomaly detection,'' \emph{Journal of API Security}, 2021.

\bibitem{sandoval2024phantom}
\BIBentryALTinterwordspacing
K.~Sandoval, ``Understanding the phantom token approach,'' July 2024, accessed: 2025-04-04. [Online]. Available: \url{https://nordicapis.com/understanding-the-phantom-token-approach/}
\BIBentrySTDinterwordspacing

\bibitem{hardt2017oauth}
\BIBentryALTinterwordspacing
D.~Hardt, J.~Dennis, and A.~P. Williams, \emph{OAuth 2 in Action}.\hskip 1em plus 0.5em minus 0.4em\relax Manning Publications, 2017, accessed: 2025-04-04. [Online]. Available: \url{https://www.manning.com/books/oauth-2-in-action}
\BIBentrySTDinterwordspacing

\bibitem{hardt2023oauth21}
D.~Hardt, J.~Bradley, T.~Lodderstedt, and N.~Sakimura, ``Oauth 2.1 authorization framework (draft),'' \url{https://datatracker.ietf.org/doc/html/draft-ietf-oauth-v2-1-08}, 2023.

\bibitem{alaca2016device}
F.~Alaca and P.~C. Van~Oorschot, ``Device fingerprinting for augmenting web authentication: Classification and analysis of methods,'' \emph{Proceedings of the IEEE}, vol. 104, no.~8, pp. 1649--1663, 2016.

\bibitem{cherif2023zero}
\BIBentryALTinterwordspacing
A.~Nait~Cherif, Y.~Achir, M.~Youssfi, M.~Elgarej, and O.~Bouattane, ``Zero-knowledge proofs and oauth 2.0 for anonymity and security in distributed systems,'' in \emph{Proceedings of the 5th International Conference on Electronics, Energy, and Measurement (ICEGC)}, E3S Web of Conferences.\hskip 1em plus 0.5em minus 0.4em\relax EDP Sciences, 2023. [Online]. Available: \url{https://www.e3s-conferences.org/articles/e3sconf/pdf/2023/106/e3sconf_icegc2023_00085.pdf}
\BIBentrySTDinterwordspacing

\bibitem{thakur2015user}
M.~A. Thakur and R.~Gaikwad, ``User identity and access management trends in it infrastructure-an overview,'' in \emph{2015 International Conference on Pervasive Computing (ICPC)}.\hskip 1em plus 0.5em minus 0.4em\relax IEEE, 2015, pp. 1--4.

\bibitem{ferry2015security}
E.~Ferry, J.~O~Raw, and K.~Curran, ``Security evaluation of the oauth 2.0 framework,'' \emph{Information \& Computer Security}, vol.~23, no.~1, pp. 73--101, 2015.

\bibitem{siriwardena2014advanced}
P.~Siriwardena, ``Advanced api security,'' \emph{Apress: New York, NY, USA}, 2014.

\bibitem{ethelbert2017json}
O.~Ethelbert, F.~F. Moghaddam, P.~Wieder, and R.~Yahyapour, ``A json token-based authentication and access management schema for cloud saas applications,'' in \emph{2017 IEEE 5th International Conference on Future Internet of Things and Cloud (FiCloud)}.\hskip 1em plus 0.5em minus 0.4em\relax IEEE, 2017, pp. 47--53.

\bibitem{rodriguez2001analysis}
P.~Rodriguez, C.~Spanner, and E.~W. Biersack, ``Analysis of web caching architectures: Hierarchical and distributed caching,'' \emph{IEEE/ACM Transactions On Networking}, vol.~9, no.~4, pp. 404--418, 2001.

\bibitem{ferraiolo1995role}
D.~Ferraiolo, J.~Cugini, D.~R. Kuhn \emph{et~al.}, ``Role-based access control (rbac): Features and motivations,'' in \emph{Proceedings of 11th annual computer security application conference}, 1995, pp. 241--48.

\bibitem{bertino2003rbac}
E.~Bertino, ``Rbac models—concepts and trends,'' \emph{Computers \& Security}, vol.~22, no.~6, pp. 511--514, 2003.

\bibitem{sandhu1994access}
R.~S. Sandhu and P.~Samarati, ``Access control: principle and practice,'' \emph{IEEE communications magazine}, vol.~32, no.~9, pp. 40--48, 1994.

\bibitem{sun2012systematically}
S.-T. Sun, K.~Hawkey, and K.~Beznosov, ``Systematically breaking and fixing openid security: Formal analysis, semi-automated empirical evaluation, and practical countermeasures,'' \emph{Computers \& Security}, vol.~31, no.~4, pp. 465--483, 2012.

\bibitem{das2012key}
S.~Das, Y.~Ohba, M.~Kanda, D.~Famolari, and S.~K. Das, ``A key management framework for ami networks in smart grid,'' \emph{IEEE Communications Magazine}, vol.~50, no.~8, pp. 30--37, 2012.

\bibitem{benson2020disco}
L.~Benson, P.~M. Grulich, S.~Zeuch, V.~Markl, and T.~Rabl, ``Disco: Efficient distributed window aggregation.'' in \emph{EDBT}, vol.~20, 2020, pp. 423--426.

\bibitem{luo2021deep}
Y.~Luo, Y.~Xiao, L.~Cheng, G.~Peng, and D.~Yao, ``Deep learning-based anomaly detection in cyber-physical systems: Progress and opportunities,'' \emph{ACM Computing Surveys (CSUR)}, vol.~54, no.~5, pp. 1--36, 2021.

\bibitem{ranjan2022user}
\BIBentryALTinterwordspacing
R.~Ranjan and S.~S. Kumar, ``User behaviour analysis using data analytics and machine learning to predict malicious user versus legitimate user,'' \emph{High-Confidence Computing}, vol.~1, p. 100034, 2022. [Online]. Available: \url{https://doi.org/10.1016/j.hcc.2021.100034}
\BIBentrySTDinterwordspacing

\bibitem{yang2016model}
\BIBentryALTinterwordspacing
R.~Yang, G.~Lee, W.~C. Lau, K.~Zhang, and P.~Hu, ``Model-based security testing: An empirical study on oauth 2.0 implementations,'' in \emph{Proceedings of the 11th ACM on Asia Conference on Computer and Communications Security}.\hskip 1em plus 0.5em minus 0.4em\relax ACM, 2016, pp. 651--662. [Online]. Available: \url{https://dl.acm.org/doi/10.1145/2897845.2897874}
\BIBentrySTDinterwordspacing

\bibitem{rahat2022cerberus}
\BIBentryALTinterwordspacing
T.~A. Rahat, Y.~Feng, and Y.~Tian, ``Cerberus: Query-driven scalable vulnerability detection in oauth service provider implementations,'' in \emph{Proceedings of the 2022 ACM SIGSAC Conference on Computer and Communications Security}.\hskip 1em plus 0.5em minus 0.4em\relax ACM, 2022, pp. 3127--3141. [Online]. Available: \url{https://dl.acm.org/doi/10.1145/3548606.3559381}
\BIBentrySTDinterwordspacing

\bibitem{ramamoorthi2024anomaly}
V.~Ramamoorthi, ``Anomaly detection and automated mitigation for microservices security with ai,'' \emph{Applied Research in Artificial Intelligence and Cloud Computing}, vol.~7, no.~6, pp. 211--222, 2024.

\bibitem{serbout2023api}
S.~Serbout, A.~El~Malki, C.~Pautasso, and U.~Zdun, ``Api rate limit adoption--a pattern collection,'' in \emph{Proceedings of the 28th European Conference on Pattern Languages of Programs}, 2023, pp. 1--20.

\bibitem{mathur2024artificial}
S.~Mathur, Y.~Hasan, D.~Bhargava, S.~Bhattacharjee, and A.~Rana, ``Artificial intelligence based predictive analytics for website performance optimization,'' in \emph{2024 7th International Conference on Contemporary Computing and Informatics (IC3I)}, vol.~7.\hskip 1em plus 0.5em minus 0.4em\relax IEEE, 2024, pp. 795--800.

\bibitem{simon2000brewer}
S.~Simon, ``Brewer’s cap theorem,'' \emph{CS341 Distributed Information Systems, University of Basel (HS2012)}, 2000.

\bibitem{chatgpt}
{OpenAI}, ``{ChatGPT} (mar 2024 version),'' 2024, \url{https://openai.com/chatgpt}.

\end{thebibliography}

\end{document}